\documentstyle[aps,twocolumn]{revtex}
\begin{document}
\draft
\title{
Electronic structure and valence band spectra of Bi$_4$Ti$_3$O$_{12}$
}
\author{A.~V.~Postnikov,\cite{*} St.~Bartkowski,
        F.~Mersch, and M.~Neumann}
\address{
Universit\"at Osnabr\"uck -- Fachbereich Physik,
D-49069 Osnabr\"uck, Germany
}
\author{E.~Z.~Kurmaev, V.~M.~Cherkashenko, S.~N.~Nemnonov,
and V.~R.~Galakhov}
\address{
Institute of Metal Physics, Russian Academy of Sciences -- Ural Division,
620219 Yekaterinburg GSP-170, Russia
}
\date{Received 12 May 1995}
\maketitle
\begin{abstract}
The x-ray photoelectron valence band spectrum
and x-ray emission valence-band spectra
(Ti$K{\beta_5}$, Ti$L{\alpha}$, O$K{\alpha}$)
of Bi$_4$Ti$_3$O$_{12}$ are presented
(analyzed in the common energy scale) and interpreted
on the basis of a band-structure calculation for an idealized
$I4/mmm$ structure of this material.
\end{abstract}

\section{Introduction}
\label{sec:intro}

Bi$_4$Ti$_3$O$_{12}$ belongs to the family of mixed bismuth
\linebreak \mbox{oxides} with the chemical formula
(Bi$_2$O$_2$)$^{2+}$(Me$_{x-1}R_x$O$_{3x+1}$)$^{2-}$
first synthesized and described by Aurivillius,\cite{aurivi}
which became subject to many subsequent
studies.\cite{subba,dorr71,tss89,rae90,frit92}
In the above formula, Me stands for a mono-, di-, or trivalent ion
or a mixture of those;
$R$ represents Ti$^{4+}$, Nb$^{5+}$, or Ta$^{5+}$; and $x$=2, 3, 4, etc.
The crystal structure of Bi$_4$Ti$_3$O$_{12}$ can be described
as a stacking of Bi$_2$O$_2^{2+}$ and Bi$_2$Ti$_3$O$_{10}^{2-}$ layers
along the pseudotetragonal c axis (See Fig.~1 of Ref.\cite{subba}).
In the Bi$_2$Ti$_3$O$_{10}^{2-}$ units, Ti ions are coordinated
by oxygen octahedra, which form linear chains,
and Bi ions occupy the interstitials outside the TiO$_6$ octahedra.
It was pointed out in Ref.\cite{subba} that
the Bi$_2$Ti$_3$O$_{10}^{2-}$ unit has some
similarity with the perovskite structure, and the height of the
perovskite-type layer sandwiched between Bi$_2$O$_2^{2+}$ layers
is equal to six Ti--O distances or to three ABO$_3$ perovskite units.

According to some empirical rules,\cite{matth,smoko}
such a type of structure with highly charged cations surrounded by
oxygen octahedra, which are linked through corners, is favorable
for the occurence of ferroelectricity, which was indeed found
in Bi$_4$Ti$_3$O$_{12}$ by Subbarao.\cite{subba} Recently, it was shown
that ferroelectric Bi$_4$Ti$_3$O$_{12}$ thin films can find
an application in electrooptic devices.\cite{jo_apl}

Another intriguing aspect of this material is its
relation to high-$T_c$ oxide superconductors.
As has been recently found,\cite{maeda,tuoi88}
one of the phases of the Bi-Ca-Sr-Cu-O oxide superconductors family
with $T_c\sim$80 K has a structure analogous to that
of Bi$_4$Ti$_3$O$_{12}$.  In connection with this,
some attempts for the search of different superconductors
related with ferroelectric Bi$_4$Ti$_3$O$_{12}$
have been undertaken.\cite{tss89}

The electronic structure and chemical bonding of Bi$_4$Ti$_3$O$_{12}$
has not been studied up to now. In the present paper we present,
to our knowledge, the first band-structure calculations
along with the first
measurements of x-ray photoelectron spectra (XPS) of the valence band
and x-ray emission Ti$L_{\alpha}$ ($2p_{3/2}-3d4s$ transition),
Ti$K_{\beta_5}$ ($1s-4p$ transition), O$K_{\alpha}$ ($1s-2p$ transition)
spectra (XES) of this compound, which give a direct information
about the total and partial density of states distribution
in the valence band.

\section{Experiment}
\label{sec:exp}

Single crystals of Bi$_4$Ti$_3$O$_{12}$ were grown using a modified
Nacken--Kyropoulus crystal growth apparatus. The temperature was measured
by a thermopile of four thermocouples PtRh6\% versus PtRh30\% (EL 18)
and a digital multimeter (sensitivity 0.1 $\mu$V). The multimeter and
power supply of the resistance heating system were controlled by
a computer.

According to the crystal growth experiments of Burton,\cite{burton}
the solution consisted of 68 mol\% Bi$_2$O$_3$ (Johnson Matthey, Grade 1),
20 mol\% B$_2$O$_3$ (Johnson Matthey, Grade 1) and
12 mol\% TiO$_2$ (Merck, Optipur). The platinum crucible
with the solution was placed on a ceramic support in
the upper part of the heating system. The crystals were grown
on small, thin (001) seeds.
At a temperature of about 1270 K, they were carefully placed
on the solution surface, where they stayed due to the surface tension.
Temperature gradients up to 50 K/cm provide strong convection
directed from the crucible wall to the center. As a result, the
growing crystal was fixed in the middle of the crucible.

The cooling rate was 0.1 K/h at the beginning and 0.5 K/h at the end
of the growth experiment. At a temperature of about 1200 K,
the crystal was lifted from the solution using a platinum
net and cooled down to room temperature at 15 K/h.
Crystals of about 12 g and a thickness of up to 3~mm were obtained.
Rectangular Bi$_4$Ti$_3$O$_{12}$ samples were cut and subsequently
polished to optical quality.

Ti~$L{\alpha}$ ($2p_{3/2}-3d~4s$ transition)
and O$K{\alpha}$ ($1s-2p$ transition) XES
were measured with a x-ray spectrometer (RSM-500)
with a diffraction grating ($N$=600 lines/mm; $R$=6 m) and
electron excitation.
In the course of XES measurements, all precausions
were taken in order to avoid the decomposition of the sample,
or its modification under the electron bombardment:
a very soft regime with a low current on the x-ray tube
($I$=0.3 mA) was taken, and the position
of the sample with respect to the electron beam was changed
for each scan.
The spectra were recorded in the first order
of reflection by a secondary electron multiplier with CsI photocathode
in an oil-free vacuum of $(1-2)\times10^{-6}$ Torr.
The entrance and exit slits of the spectrometer were
20 $\mu$m wide, which gave an instrumental resolution Ti~$L$
and O~$K{\alpha}$ XES of about 0.9 and 1.2 eV, respectively.
The x-ray tube was operated at $V$=4.6 keV.

Ti~$K{\beta_5}$ XES ($1s-4p$ transition)
was measured using fluorescent excitation
with a Johan-type x-ray tube spectrometer
with a position sensitive detector.\cite{dolgih}
A ($10{\bar1}1$) plane of quartz curved to $R$=2000 mm
was used as a crystal analyzer.
The primary Cu~$K$ radiation was used for the excitation of
the Ti~$K{\beta_5}$ spectra. The energy resolution was 0.3 eV. The sealed
x-ray tube with a Cu anode was operated at $V$=35 keV, $I$=50 mA.
The time of accumulation of the Ti~$K{\beta_5}$ spectra was about 50 h.

X-ray photoelectron and core-level spectra of Bi$_4$Ti$_3$O$_{12}$
were measured using an ESCA spectrometer of Perkin-Elmer
(PHI 5600 ci, monochromatized Al~$K{\alpha}$ radiation, FWHM=0.3 eV).
The resolution of the concentric hemispherical analyzer
was adjusted to less than $\Delta{E}=0.2$ eV.
A single crystal of Bi$_4$Ti$_3$O$_{12}$ was cleaved in high vacuum
prior to the XPS measurements. The spectra were calibrated
using the C~$1s$ line of small carbon contaminations
of the surface as a reference level
($E$=285.0 eV). A survey XPS from the valence-band region
is shown in the upper panel of Fig.~\ref{vb-2}.

The unmarked feature at a binding energy of 23.5 eV
is due to partial decomposition of the sample
in the course of measurements that resulted in the formation
of metallic Bi at the sample surface. In order to avoid
the charging, it was necessary to irradiate the sample with
low kinetic energy electrons ($E_{kin}\leq$100 eV).
In the lower panel of Fig.~\ref{vb-2}, the evolution of the spectrum
with the measurement time is shown.
It has not been investigated whether the decomposition
of the sample was caused by electrons or x-ray photons.

Based on the measured binding energies of the core levels from XPS
(531.50 eV for O~$1s$, 459.80 eV for Ti~$2p_{3/2}$),
it is possible to align the x-ray photoemission
spectrum of the valence band and the individual x-ray emission
spectra to a common energy scale with respect to the Fermi level.
Since for dielectric materials the error of several eV
in relative positions of the spectra
due to charging effects is difficult to overcome,
we had to displace the spectra slightly
in order to match the positions of the valence band
in the XPS and in the XES of components.
More detailed structure of the valence band XPS and XES
on a common energy scale is shown in Fig.~\ref{vb-1}.

\section{Structure model and electronic structure calculations}
\label{sec:calc}

The crystal structure of Bi$_4$Ti$_3$O$_{12}$ has been determined
from x-ray and neutron diffraction experiments
by Dorrian {\it et~al.}\cite{dorr71}
and was afterwards refined by David Rae {\it et~al.}.\cite{rae90}
The material was found to be a displacive ferroelectric
with polar orthorombic structure, with four formula units
per unit cell and the lattice parameters
$a$=5.45~\AA, $b$=5.41~\AA, $c$=32.83~\AA.
The space group is $B2cb$ according to Ref.~\cite{dorr71}
and $B1a1$ as found in Ref.~\cite{rae90}.
Any of these orthorombic structures is a commensurate modulation
of the nonpolar orthorombic parent $Fmmm$ structure,
which is derived from the idealized tetragonal $I4/mmm$ structure.
Bi$_4$Ti$_3$O$_{12}$ exists in a $Fmmm$ structure above
the ferroelectric transition temperature of 675~$^0$C.

We performed {\it ab initio} band-structure calculation
within the local density approximation (LDA), with the exchange-correlation
potential parametrization according to von Barth and Hedin\cite{vbh}
and gradient corrections as proposed by Langreth and Mehl.\cite{lm}
In spite of some argument that the multielectron correlation
effects beyond the LDA play a role in the formation
of electronic sructure of titanium oxides\cite{kotani},
the only noticeable effect of correlations, if any,
seems to be confined to the core-level region\cite{correl}.
Therefore, one may expect the LDA to provide
an adequate description of the occupied electronic states
within the valence band.

Specifically, we used the
tight-binding linear muffin-tin orbital method (TB-LMTO)
in the atomic sphere approximation (ASA)\cite{tblmto},
which was proven to be quite accurate and efficient
in many applications. Even with this method,
the electronic structure calculation of the real orthorombic
phase with 76 atoms per unit cell poses a problem.
It is even more complicated by the fact that the crystal structure
is rather open, apart from the chains of Ti-O octahedra, which are
intersected by warped Bi-O-Bi layers. In order to provide
an adequate description of the potential over the unit cell,
the structure has to be packed with additional empty spheres,
which increase the basis set and the complexity of the
computational problem even further.

In order to keep the computational effort
in our band structure calculations manageable
and to simplify the analysis of underlying trends
in the electronic structure of the material,
we prefered to perform calculations
for the idealized $I4/mmm$ crystal structure.
It is a bodycentered tetragonal structure
for which we used lattice parameters $a$=7.2558 a.u.
(=$\sqrt{ab/2}$ of the orthorombic structure)
and $c/a$=8.55. We used the atomic positions
listed in Table 6 (Model 3) of Ref.\cite{rae90}
and averaged them over sites that become equivalent
in the $I4/mmm$ structure. The resulting positions
of atoms and empty spheres ($E1$ to $E4$) are listed
in Table \ref{bitio:struc}.

The positions of the atoms may easily be seen
from the charge density plots
in two different planes cutting the unit cell, as obtained
from self-consistent TB-LMTO calculation (Fig.~\ref{rho}).
Due to the inversion symmetry,
only half of the unit cell is shown.
Moreover, a $[\frac{1}{2}\frac{1}{2}\frac{1}{2}]$ translation
maps the atoms and the charge density pattern onto itself.
In the actual orthorombic structure,
O4-Ti2-O3-Ti1-O3-Ti2-O4 chains of joined TiO$_6$ octahedra
are rotated around the $c$ axis by up to 7.5$^0$
(in the opposite direction for adjacent chains),
shifted in the plane normal to $c$, and a tilting
up to 6.5$^0$ appears for octahedra within each chain.
These distortions, which are well described
in Ref.\cite{rae90}, are not negligible
but they are expected to have only a moderate effect on the basic
band-structure properties, because the atomic coordination
and distances between near neighbors are essentially
preserved. Since in the present work we are mostly concerned with
spectroscopic information that is accumulated
from crystallographically inequivalent atoms anyway
and do not discuss the aspects of ferroelectric behavior,
our structure model used in the calculations
seems to be justified.

The choice of the radii of space-filling atomic spheres
for a LMTO-ASA calculation is not unique for compounds, especially
for those including several types of atoms with different
electronegativity.
Our choice was based on a compromise between
possibly good matching of potentials
at the sphere boundaries for all types of atoms
and possibly low spheres overlap which we managed
to keep below 30\% of all interatomic distances.
The final values of the atomic sphere radii $S$ together with
the boundary values of the potential, adjusted in the course
of iterations, are listed in Table \ref{bitio:char}
along with the self-consistent partial charges inside the spheres.
There is not much sense to consider the net charge ${\Delta}Q$ within
each atomic sphere, which is listed in the last column
of Table \ref{bitio:char}, as direct reference to
a charge state of the atoms in the compound, because this
property of course depends on the definition and may vary
considerably as based on the different estimates. It is reasonable,
however, to analyze the changes of ${\Delta}Q$ from one
crystallographically inequivalent position to another
within the same chemical species, because this gives
clear indication of the charge transfer.
The partial densities of states (DOS) of the components
as obtained from TB-LMTO calculations
using the tetrahedron method with 140 {\bf k} points
in the irreducible part of the Brillouin zone are shown
in Figs.~\ref{ti}, \ref{o} and \ref{bi}.

\section{Results and discussion}
\label{sec:disc}

As is seen from Fig.~\ref{rho}, the Ti atoms in Bi$_4$Ti$_3$O$_{12}$
are surrounded by almost perfect O$_6$ octahedra.
With Bi1 and Bi2 atoms occupying the pores formed by 12 oxygen atoms
of adjacent octahedra,
a perovskite-type sequence of layers is formed:
(Bi2-O4); (O5-Ti2-O5); (Bi1-O3); (O1-Ti1-O1);
(Bi1-O3); (O5-Ti2-O5); (Bi2-O4).
The corresponding repeated perovskite structure
with the chemical formula BiTiO$_3$ may, however, not be stable
due to electrostatic considerations. The system
gets stabilized by interchanging three perovskite-type layers
with an additional O2 plane. This is accompanied by
considerable warping of (Bi2-O4) planes and
a noticeable [001] displacement of the Ti2 atoms from the middle
of O$_6$ octahedra surrounding them.

According to our calculations, the $3d$ DOS at the Ti1 site
(Fig.~\ref{ti}) is rather similar to that known for
perovskite materials as BaTiO$_3$ and PbTiO$_3$
(see, e.g., Ref.\cite{cohen1,cohen2}) -- the Ti$3d$ states
are concentrated at the bottom of the broad (about 5 eV)
valence band. Correspondingly, the O~$2p$ DOS
at the O1, O3 and O5 sites (Fig.~\ref{o}) are rather typical
for oxygen in perovskites, with a pronounced maximum
near the top of the valence band where O$2p$ states
are nonbonding. The nonsphericity
of the charge density around the displaced Ti2 atom
in its oxygen cage is quite similar to that around Ti in
tetragonally distorted BaTiO$_3$ (see Fig.~4 of Ref.~\cite{cohen1}).

A considerable difference in the ${\Delta}Q$ values
at the Ti1 and Ti2 sites is only partially due to a small
differences in the atomic sphere sizes, but
comes mostly from a strong dynamic charge transfer
along the Ti2-O4 bond that occurs as this bond length
is varied. This effect is well known for Ti-based perovskite
systems and is the driving force for tetragonal ferroelectric
instability in BaTiO$_3$ and PbTiO$_3$. Quantitatively,
this charge transfer or dynamic polarization of
the Ti-O bond is characterized by Born effective charges,
which were calculated by Zhong {\it et al.}\cite{zhong}
to be ${\sim}+$7 at the Ti site and ${\sim}-$5 at the O site
(as displaced towards Ti) in BaTiO$_3$ and PbTiO$_3$,
i.e., very different from nominal ionic valence values.

The reason why the O4 atoms aquire positive net charge
in our calculations rather than become more electronegative
as Ti2 comes closer, is that the charge distribution
at the O4 sites is far from being spherical symmetric and well
localized. O4 terminates the chain of TiO$_6$ octahedra
and is separated from the Bi2 atom by an interstitial where we
put an empty sphere $E4$. The charge transferred from Ti2
towards O4 is, therefore, partially attributed to $E4$
(and also to $E1$) empty spheres in our calculation.
The energetics of the $2s$ states (Fig.~\ref{o})
reveals that the effective potential is indeed
less attractive for electrons at the O4 as compared
to all other oxygen positions, thus indicating
higher electronegativity, due to the Ti2$\rightarrow$O4
charge transfer. Correspondingly, the center of gravity
of the O~$2p$ states is shifted to higher energies
at the O4 site. Since the valence band is essentially
formed by hybridized O~$2p$, Ti~$3d$ and Bi~$6p$ states,
the presence of only one Ti neighbor and the remoteness
of the Bi neighbors to the O4 site isolates the O~$2s$ states
to a narrow flat band and reduces the effective width
of the O4-$2p$ partial DOS.

The O2 sites, in difference to the other oxygen sites,
which belong to the octahedra cage,
form a basal plane for empty pyramides, or
semioctahedra, with the apical points (the
by O4 atoms) protruding up or down from the O2 plane
in a chessboard configuration. The O2 atoms mediate
the interaction between two Bi2 layers, and
the $2p$ states of O2 exhibit noticeable hybridization with
the $6s$ and $6p$ states of Bi2.

Summing up over all oxygen sites, the O~$2p$ partial DOS
is asymmetric, with states near the top of the valence band
dominating in the perovskite layers, and narrow band
related to O4 sites giving the main contribution again at the
upper half of the valence band. This state distribution
is well revealed in the O~$K{\alpha}$ x-ray emission
spectrum (Fig.~\ref{o}, lower panel), which probes the
$1s-2p$ transition and thus is expected to scan the occupied part
of the O~$2p$ partial DOS.

The x-ray spectra of Ti are shown in Fig.~\ref{ti}.
The Ti~$K{\beta}$ spectrum probes the occupied part of the Ti~$4p$ DOS.
The lower of two pronounced peaks separated by $\sim$15 eV
reveals the admixture of Ti~$4p$ states to the O~$2s$ band.
This energy separation is typical for all oxides
and is well reproduced in the band structure calculation.

The Ti~$L$ spectrum consists of two subbands
($L{\alpha}$ and $L{\beta}$, see Fig.~\ref{vb-1})
which correspond to x-ray transitions
with a valence band electron filling the core vacancy
in the $2p_{3/2}$ or $2p_{1/2}$ level, correspondingly.
Since the Ti~$3d$ DOS in the valence band dominate those
of the Ti~$4s$ as is seen in Fig.~\ref{ti} (note the different scales
for $3d$ and $4s$ DOS),
both subbands essentially reveal the same Ti~$3d$ DOS distribution,
modulated by a slightly different energy dependence of the dipole
transition matrix elements for $2p_{1/2}-3d$
and $2p_{3/2}-3d$ processes, and somehow overlap.
In order to enable the comparison with
the calculated DOS, we cut the experimental Ti$L{\alpha,\beta}$
spectrum near the minimum and shifted the upper part
downwards by 5.5 eV, that is the value
of the Ti~$2p_{1/2}$--Ti~$2p_{3/2}$
splitting in TiO$_2$.\cite{ti2p} Thus overimposed spectra,
when compared with the calculated Ti~$3d$ DOS (Fig.~\ref{ti},
right panel), reveal all essential features of
the occupied part of the Ti~$3d$ band.

According to our band-structure calculation,
the two inequivalent types of Bi sites have quite different
partial DOS, which reveals the differences in their
crystallographic environment.
The $6s$ states of Bi1 atoms, which belong to the perovskitelike
''BiTiO$_3$'' fragment of the crystal lattice,
are filled and separated from the valence band by a gap of about 3 eV.
The split-off Bi~$6s$ band of about 1 eV width is slightly hybridized
with O~$2p$ states at the neighboring O1, O3 and O5 sites
(see Fig.~\ref{o}). Some small fraction of the Bi~$6s$ states
is moreover admixed to the low-lying O~$2s$ states
and to the higher states in the valence band.
The partial DOS of Bi1 closely resembles that of Pb
in PbTiO$_3$ (a tetragonally distorted perovskite,
see Fig.~3 of Ref.~\cite{cohen2}),
with the only difference that the $6s$ states of Bi,
which have one electron more than Pb,
are deeper (by $\sim$2 eV with respect to the
valence band), more localized and exhibit less admixture
to the states in the valence band.

The $6p$ states of Bi1 participate mostly
in the valence band, forming a pronounced bonding
peak due to the hybridization with O~$2p$ states
at ${\sim}-$5.2 eV. This bonding state is visible
in the O~$2p$ DOS at O1, O3 and to a smaller extent at O5 sites.
There is no essential charge transfer from the Bi1 to
the neighboring O atoms. On the contrary, the large atomic sphere
at the Bi1 site accumulates some extra charge
(due to charge transfer from Ti) as compared to the free Bi atom.

In contrast to this, Bi atoms, which belong to Bi2--O2
warped planes, exhibit a considerable loss of charge
due to a more polar type of the underlying bonding
than that in the perovskite-like fragments.
The Bi2-$6s$ states are not fully occupied
and participate in the chemical bonding. The partial DOS
at the Bi2 site, with a $\sim$3 eV broad Bi~$6s$ band
and a non-negligible $6s$ contribution just below
the Fermi level, has some similarity with that
calculated for $\delta$-Bi$_2$O$_3$ in Ref.\cite{zhukov}.

A comparison of the valence band XPS with the calculated
total DOS per unit cell is shown in Fig.~\ref{tot}.
The general width and the position of the maximum
of the valence band formed mostly by Ti~$3d$,
Bi~$6p$ and O~$2p$ states are found in the calculation
to agree well with the spectroscopic data.
The position of the next distinct peak
(at about $-$10 eV in Fig.~\ref{tot}), which reveals
the Bi~$6s$, states is again correctly predicted
in the LMTO calculation. It seems possible that
the calculation overestimates the splitting
of the Bi~$6s$ peak (into two energetically separated parts
related to the Bi1 and Bi2 sites) that would otherwise
be detected in XPS at a given energy resolution.
This discrepancy may be probably eliminated
in a calculation considering a more exact structure
model, where the relaxation of the TiO$_6$ fragments
makes the potentials at inequivalent Bi sites
more similar.

The XPS peak at about $-$19 eV
in Fig.~\ref{tot}, which reveals the contribution
from the O~$2s$ states, lies approximately 2.5 eV below
the position of the O~$2s$ states according to
the LMTO calculation. This discrepancy
is typical for LDA-based electronic structure calculations of oxides,
as compared to XPS results, and is seemingly due to the effect
of the hole relaxation, which effectively increases
the binding energy of an electron leaving a comparatively
localized state such as Oi~$2s$. This effect remains beyond
the LDA, but may be correctly accounted for by more
accurate treatment of correlation effects, e.g., in a $GW$ approximation,
as was shown for MgO in Ref.~\cite{mgo-gw}.
This effect is smaller for less localized states
in the valence band.

Another discrepancy between the calculation and XPS is that
the split-off narrow peak associated with O~$2s$ states
of O4 atoms is not resolved in the measured spectrum.
It is not clear at the moment whether the realistic
geometry with relaxed TiO$_6$ octahedra smears out
the difference between O4 and other oxygen sites.
The effect of the O~$2s$ holes, which should be stronger for
more localized O4-$2s$ states, also tends to
increase the binding energy for the latter and thus
to lower their separation from other O~$2s$ states
in the XPS.
It may also well be the case that two marked discrepancies
from the experimental spectra (the splittings
of the Bi~$6s$ and O~$2s$ bands) are due to the crudeness
of ASA when applied to a quite open structure of
Bi$_4$Ti$_3$O$_{12}$, that can be checked in
subsequent calculations not using any shape approximation
for the potential.

\section{Conclusion}
\label{sec:conclu}

In the present paper, we present an analysis of the electronic structure
of Bi$_4$Ti$_3$O$_{12}$ by different spectroscopic techniques
(XPS and x-ray emission spectroscopy), combined with an
{\it ab~initio} band-structure calculation.
It was studied how the observed valence-band spectra
are formed by contributions from crystallographically inequivalent
types of Ti (two species), Bi (two species) and O (five species) atoms
in the idealized $I4/mmm$ structure.
The main difference in the charge state and the type of chemical
bonding is between those species of Bi and O,
which belong to perovskitelike fragments, on the one hand,
and to intermediate Bi--O planes on the other hand.
The electronic structure related to perovskitelike fragments
has many common features with those of perovskite-type compounds
BaTiO$_3$ and PbTiO$_3$, whereas the intermediate Bi--O planes
reveal some similarity with the bismuth oxides.

\acknowledgements

Financial support by the Deutsche Forschungsgemeinschaft (SFB~225),
the NATO (grant No. HTECH.LG940861), and
the Russian Foundation for Fundamental Research (grant No. 94-03-08040)
is gratefully acknowledged.
One of us (E.Z.K) wants to thank for the kind hospitality
at the University of Osnabr\"uck during his stay.

\begin{figure}
\caption{
Upper panel: x-ray photoelectron spectrum of Bi$_4$Ti$_3$O$_{12}$.
Lower panel: Bi~$5d$-related part of the spectrum
(from different sample)
immediately after the fracture (dots)
and after 24 h of irradiation (solid line).
}
\label{vb-2}
\end{figure}

\begin{figure}
\caption{
x-ray photoelectron spectrum and x-ray emission
O~$K{\alpha}$, Ti~$L{\alpha}$, Ti~$L{\beta}$ and Ti~$K{\beta_5}$
spectra of Bi$_4$Ti$_3$O$_{12}$.
}
\label{vb-1}
\end{figure}

\begin{figure}
\caption{
Charge density contour plots in the (010) and (110) planes
of Bi$_4$Ti$_3$O$_{12}$ as calculated by TB-LMTO method.
}
\label{rho}
\end{figure}

\begin{figure}
\caption{
Calculated partial DOS at Ti1 and Ti2 sites
and Ti x-ray emission spectra of Bi$_4$Ti$_3$O$_{12}$.
Left panel: 4$p$ (above); total Ti~4$p$ DOS per unit cell (below);
Ti~$K{\beta}$ spectrum (dots, below).
Right panel: 4$s$ (above, dashed line, left scale);
3$d$ (above, solid line, right scale);
total Ti~3$d$ DOS per unit cell (below);
Ti~$L{\alpha}$, $L{\beta}$ spectra (dots, below).
}
\label{ti}
\end{figure}

\begin{figure}
\caption{
Calculated local DOS at five inequivalent oxygen sites
in Bi$_4$Ti$_3$O$_{12}$ (upper panels);
total O~2$p$ DOS per unit cell (lower panel, solid line)
and O~$K{\alpha}$ emission spectrum (dots).
}
\label{o}
\end{figure}

\begin{figure}
\caption{
Calculated 6$s$ and 6$p$ partial DOS at Bi1 and Bi2 sites
of Bi$_4$Ti$_3$O$_{12}$.
}
\label{bi}
\end{figure}

\begin{figure}
\caption{
x-ray photoelectron spectrum of the valence band
(dots) and calculated total DOS of Bi$_4$Ti$_3$O$_{12}$
(solid line). Note the change of scale for XPS
at $-$13 eV.
}
\label{tot}
\end{figure}

\begin{table}
\caption{
Atomic coordinates in the $I4/mmm$ structure.
}
\label{bitio:struc}
\begin{tabular}{ccccc}
Site & Wyckoff notation & $x$ & $y$ & $z$ \\
\hline
Bi1  & 2$e$ & 0             & 0      & 0.0668         \\
Bi2  & 2$e$ & 0             & 0      & 0.2113         \\
Ti1  & 1$b$ & 0             & 0      & $\frac{1}{2}$  \\
Ti2  & 2$e$ & 0             & 0      & 0.3713         \\
O1   & 2$c$ & $\frac{1}{2}$ & 0      & 0              \\
O2   & 2$d$ & $\frac{1}{2}$ & 0      & $\frac{1}{4}$  \\
O3   & 2$e$ & 0             & 0      & 0.4410         \\
O4   & 2$e$ & 0             & 0      & 0.3185         \\
O5   & 4$g$ & $\frac{1}{2}$ & 0      & 0.1175         \\
$E1$ & 8$m$ & 0.2435        & 0.2435 & 0.1601         \\
$E2$ & 4$g$ & $\frac{1}{2}$ & 0.2435 & 0.1824         \\
$E3$ & 2$e$ & 0             & 0      & 0.1383         \\
$E4$ & 2$e$ & 0             & 0      & 0.2740         \\
\end{tabular}
\end{table}

\begin{table}
\caption{
Atomic sphere radii $S$ used in the calculation;
potential values at the sphere boundaries $V(S)$;
partial $s$, $p$, and $d+f$ electron occupation numbers
and net electrostatic charges ${\Delta}Q$ within atomic spheres.
}
\label{bitio:char}
\begin{tabular}{ccccccc}
Site & $S$, a.u. & $V(S)$, Ry & $Q_s$ & $Q_p$ & $Q_{d+f}$ & ${\Delta}Q$ \\
\hline
Bi1  & 3.6 & $-$0.739 & 2.067 & 1.921 & 1.281 & $-$0.269 \\
Bi2  & 3.2 & $-$0.708 & 1.800 & 1.193 & 0.757 & $+$1.250 \\
Ti1  & 2.5 & $-$0.903 & 0.375 & 0.726 & 1.945 & $+$0.954 \\
Ti2  & 2.3 & $-$0.885 & 0.224 & 0.397 & 1.549 & $+$1.830 \\
O1   & 2.0 & $-$0.810 & 1.740 & 4.465 & 0.030 & $-$0.235 \\
O2   & 2.1 & $-$0.727 & 1.724 & 4.477 & 0.031 & $-$0.232 \\
O3   & 2.0 & $-$0.799 & 1.741 & 4.420 & 0.027 & $-$0.188 \\
O4   & 2.0 & $-$0.758 & 1.653 & 4.180 & 0.016 & $+$0.151 \\
O5   & 2.1 & $-$0.737 & 1.770 & 4.443 & 0.030 & $-$0.243 \\
$E1$ & 1.8 & $-$0.707 & 0.179 & 0.143 & 0.058 & $-$0.380 \\
$E2$ & 1.7 & $-$0.580 & 0.093 & 0.064 & 0.018 & $-$0.175 \\
$E3$ & 1.7 & $-$0.515 & 0.063 & 0.047 & 0.014 & $-$0.124 \\
$E4$ & 1.6 & $-$0.722 & 0.169 & 0.107 & 0.034 & $-$0.310 \\
\end{tabular}
\end{table}


\begin{references}
\bibitem{correl} The low-intensity Ti~$2p$ satellites
		 observed in the XPS of TiO$_2$ seem to be
		 mostly due to inelastic loss processes,
		 but the correlation effects may well play
		 a certain role in the satellite formation.
		 These aspects are now specially investigated
		 for TiO$_2$ and will be reported elsewhere;
		 in Bi$_4$Ti$_3$O$_{12}$,
		 no clear Ti~$2p$ satellite structure was observed,
		 because of the overlap with Bi~$4d$ states.
\bibitem{ti2p}   The Ti~$2p_{1/2}$--Ti~$2p_{3/2}$ splitting in
		 Bi$_4$Ti$_3$O$_{12}$ is close to this value;
		 more exact evaluation is hindered by the
		 overlap of Ti$2p_{1/2}$ and Bi$3d_{3/2}$ levels
		 in the x-ray photoemission spectrum.
\end{references}
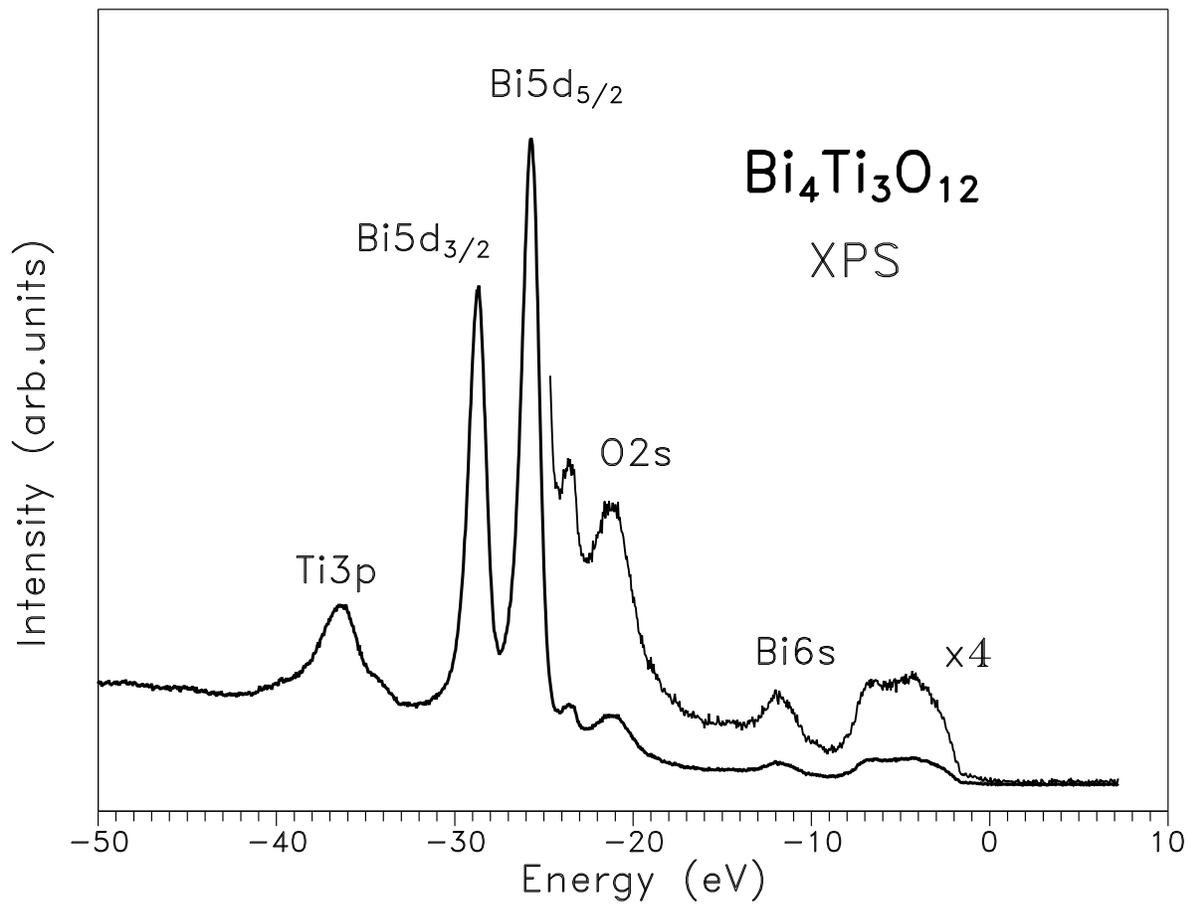
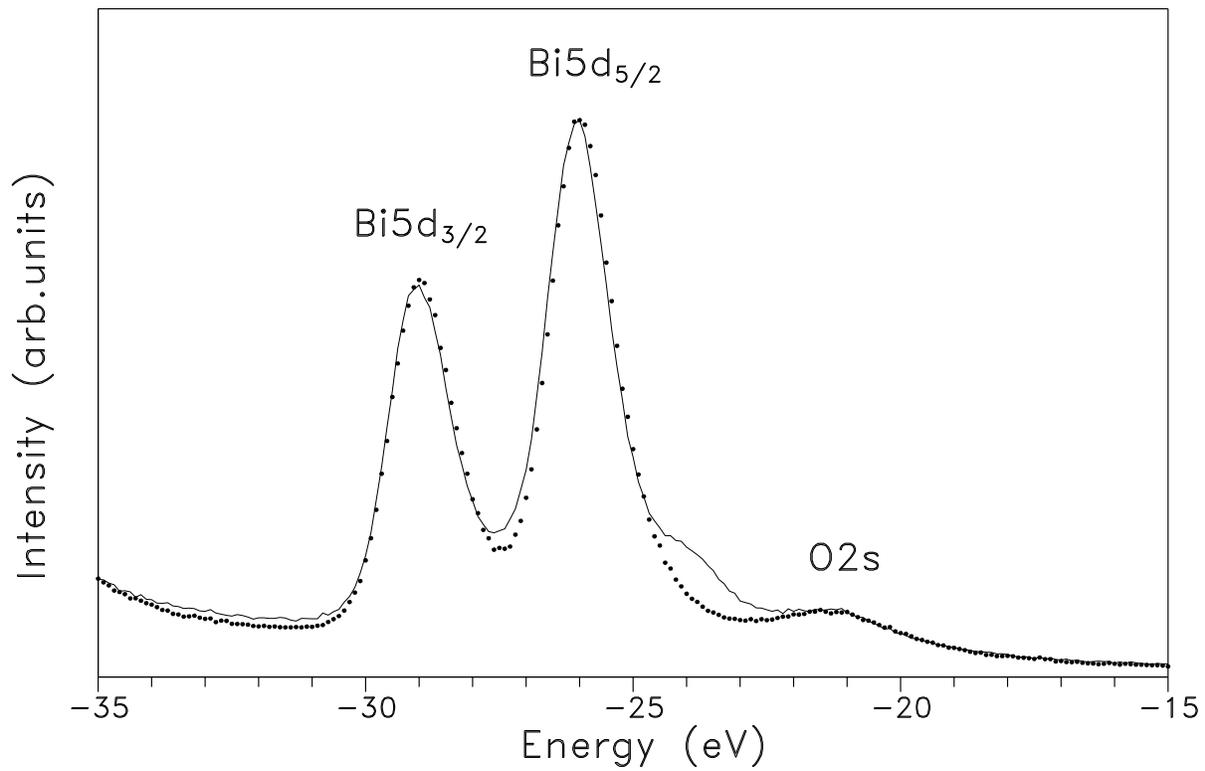
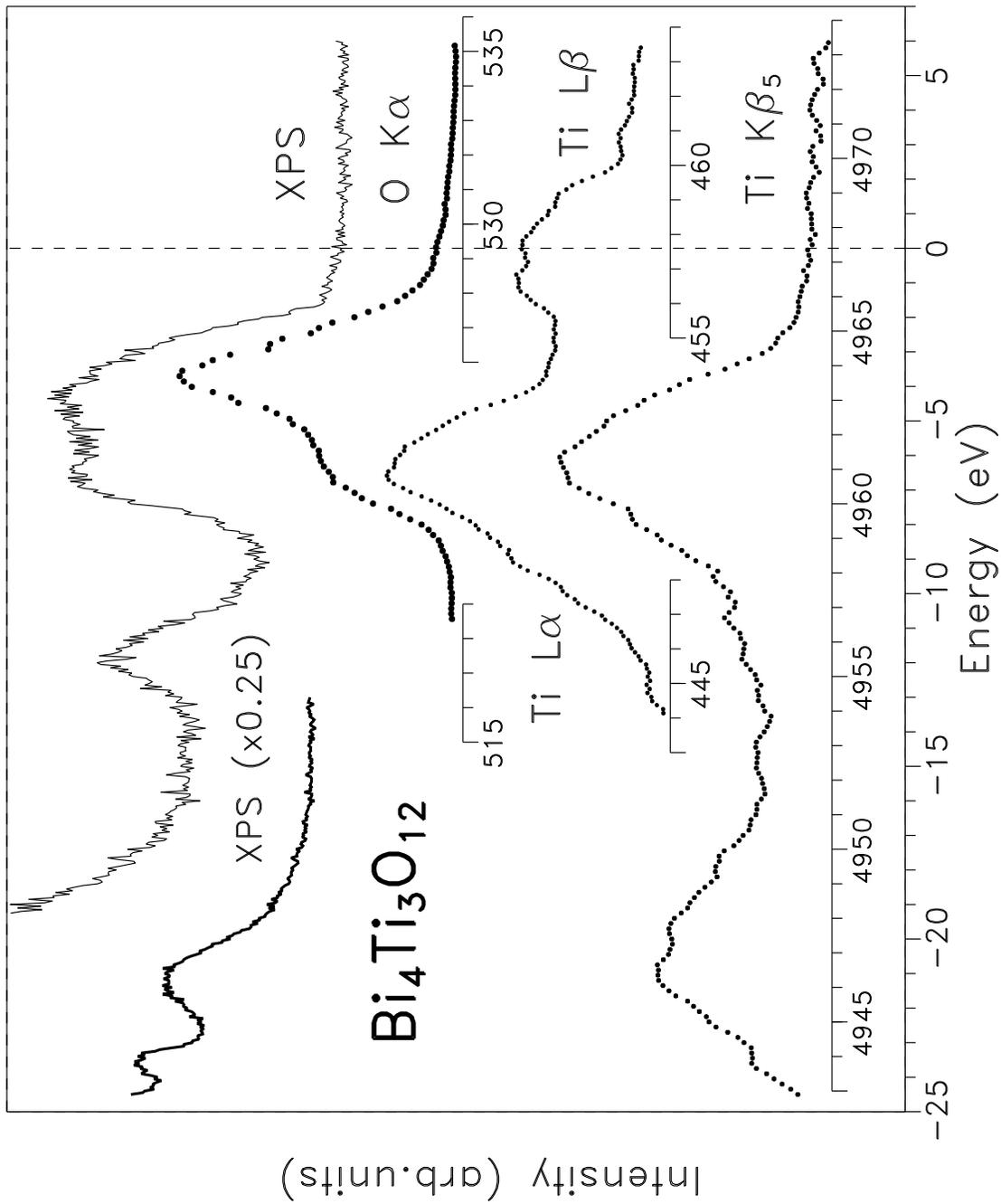
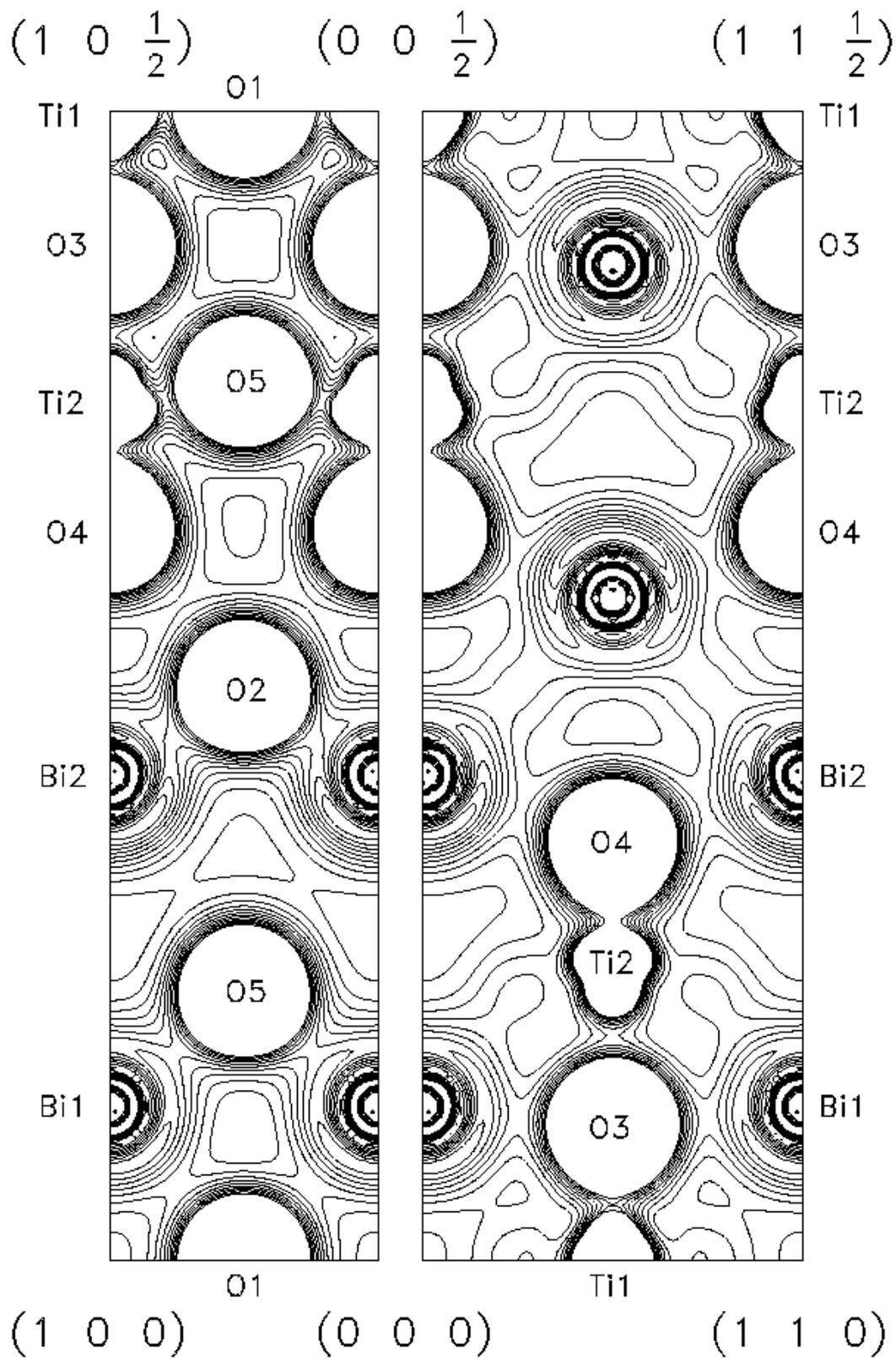
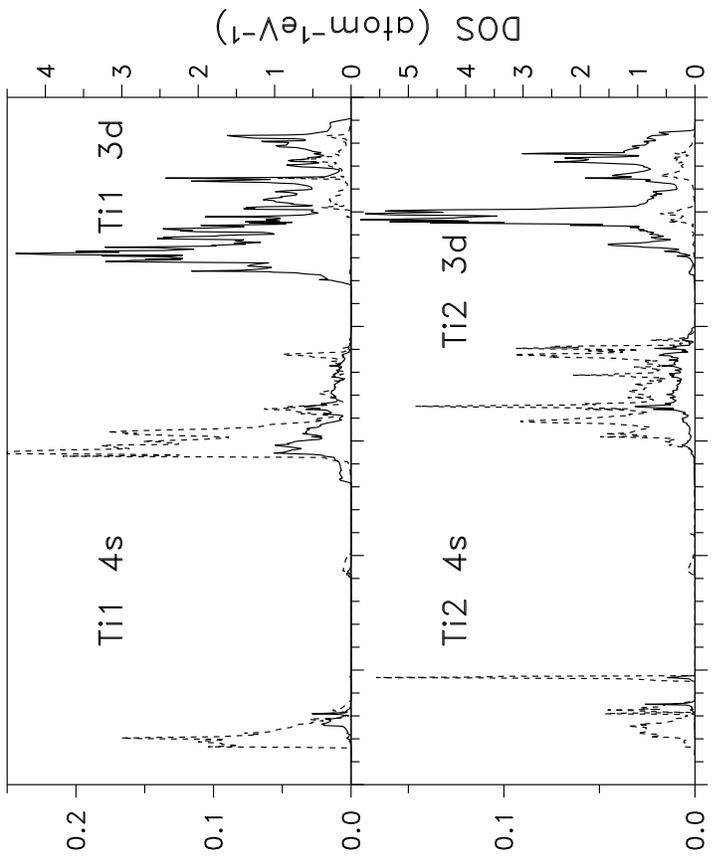
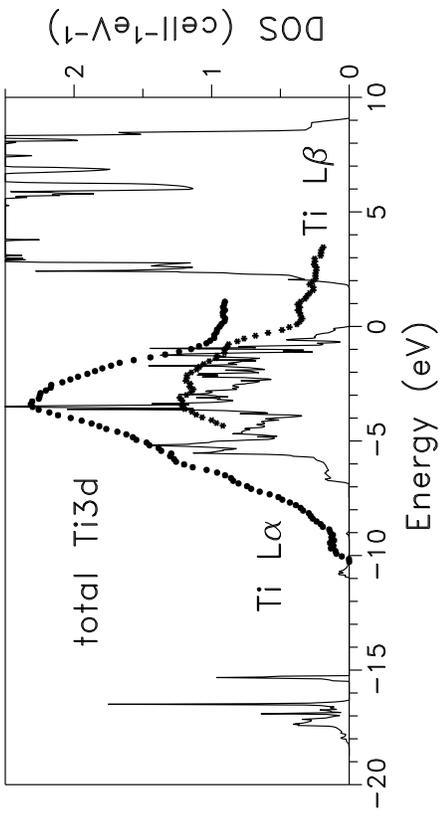
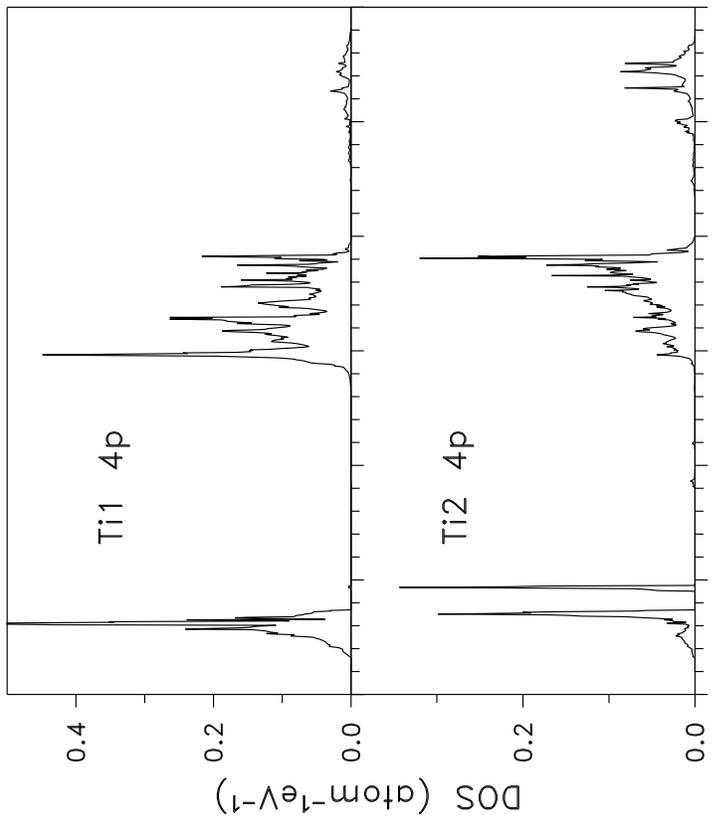
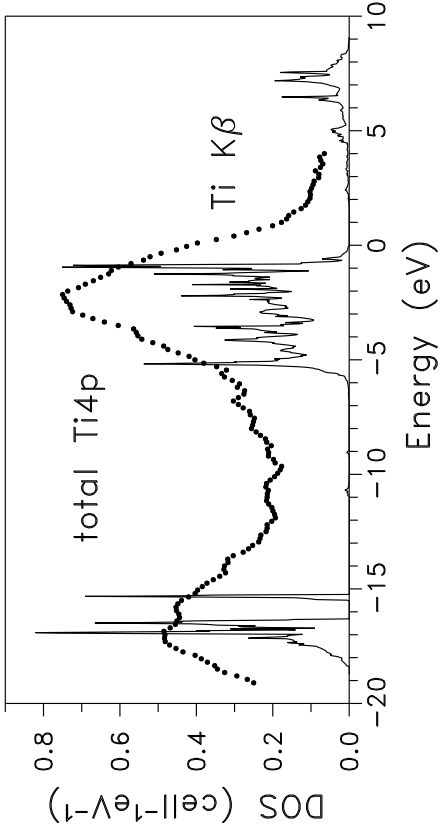
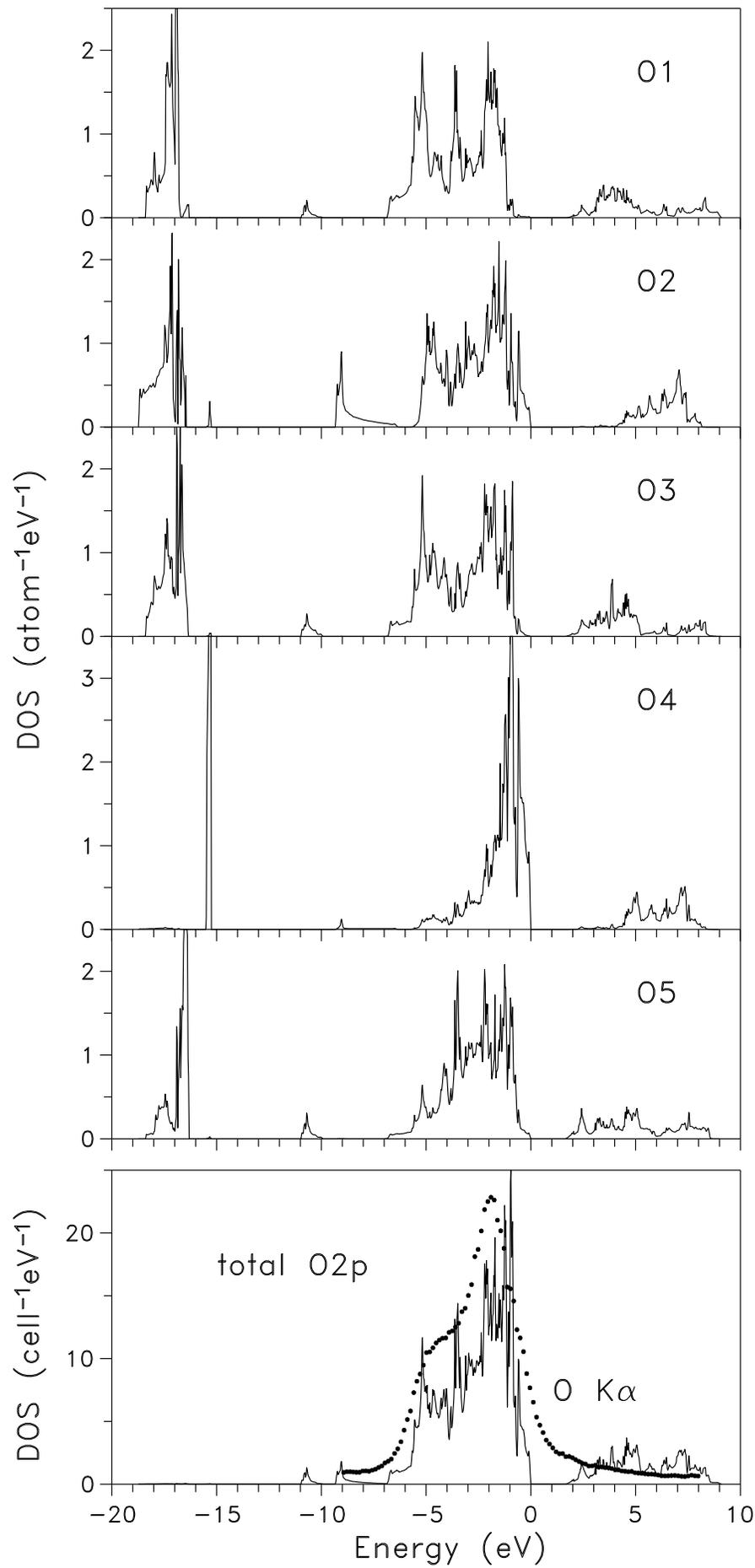
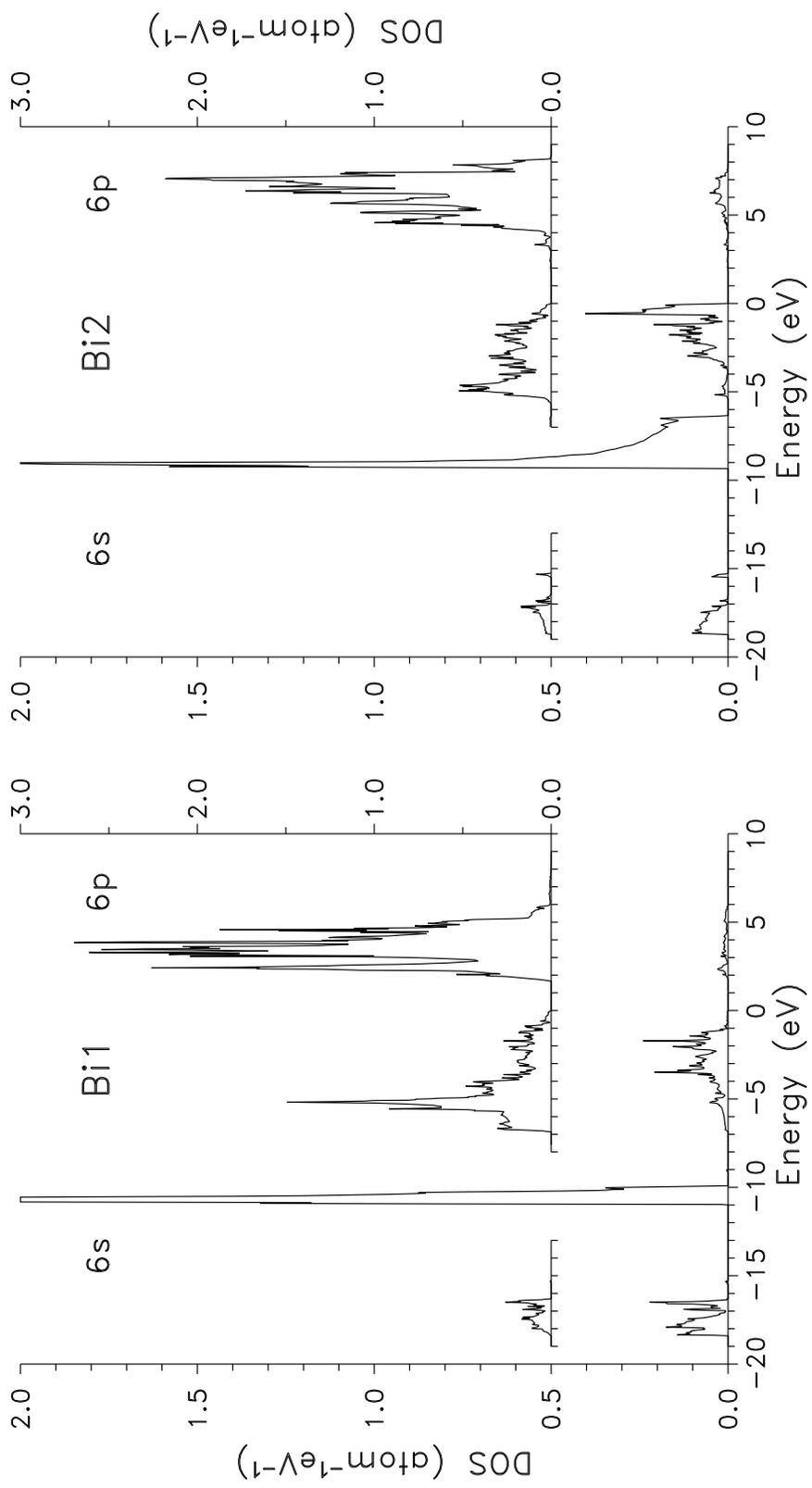
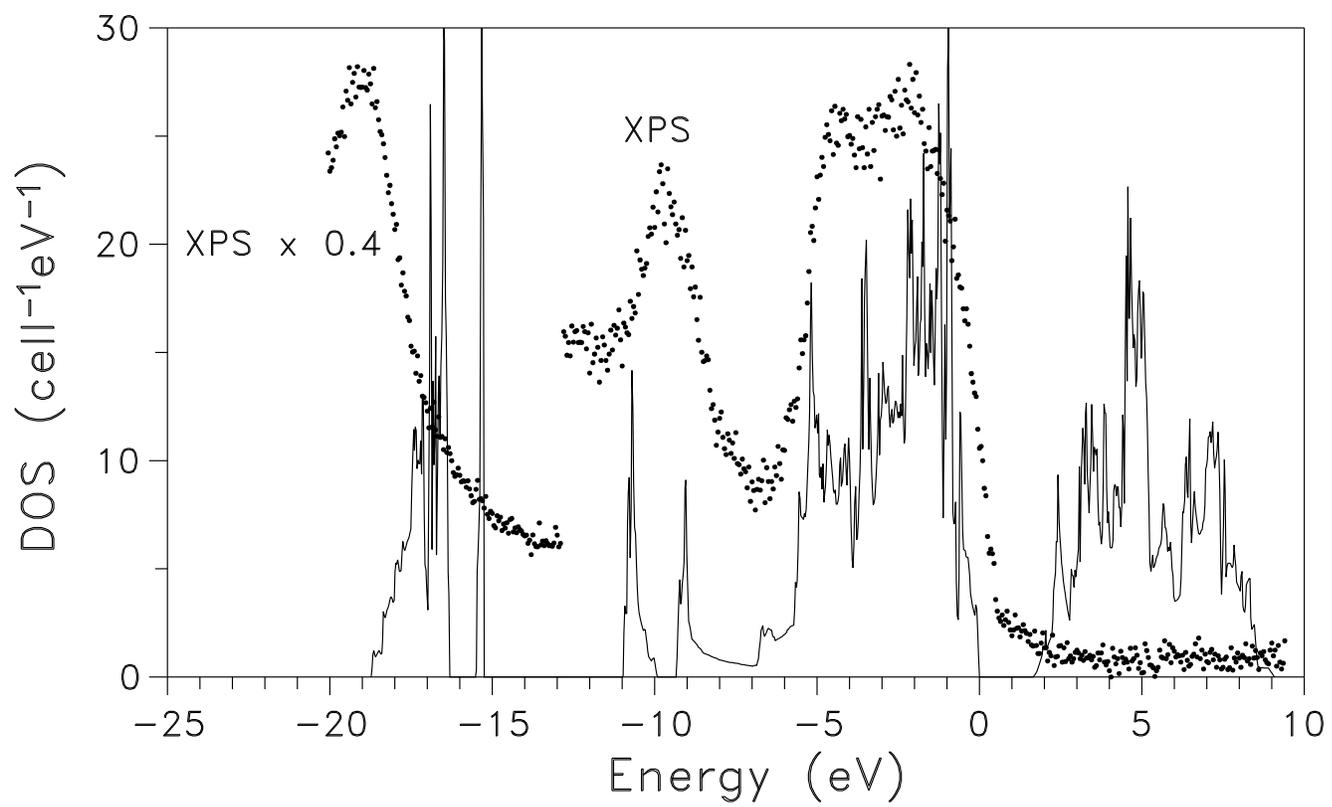
\end{document}